\documentclass[aps,prd,twocolumn,preprintnumbers,nofootinbib,amsmath,amssymb,superscriptaddress,10pt]{revtex4-2}

\usepackage{graphicx}
\usepackage{dcolumn}
\usepackage{bm}
\usepackage{amsmath}
\usepackage{braket}
\usepackage{slashed}    
\usepackage{epstopdf}
\usepackage{placeins}
\usepackage{multirow}
\usepackage{makecell}
\usepackage[shortlabels]{enumitem}
\usepackage{xcolor}
\usepackage{multirow}
\usepackage{makecell}
\usepackage{slashed}
\usepackage[hidelinks]{hyperref}
\usepackage{cleveref}

\newcommand{\beq}{\begin{eqnarray}}
\newcommand{\eeq}{\end{eqnarray}}
\newcommand{\be}{\begin{equation}}
\newcommand{\ee}{\end{equation}}
\newcommand{\beqnn}{\begin{eqnarray*}}
\newcommand{\eeqnn}{\end{eqnarray*}}
\newcommand{\benn}{\begin{equation*}}
\newcommand{\eenn}{\end{equation*}}

\newcommand{\Tr}{\ensuremath{\mathrm{Tr}}}

\newcommand{\SU}{\ensuremath{\mathrm{SU}}}

\newcommand{\MS}{\ensuremath{\overline{\mathrm{MS}}}}

\renewcommand{\P}{\ensuremath{\mathrm{P}}}
\renewcommand{\S}{\ensuremath{\mathrm{S}}}
\newcommand{\A}{\ensuremath{\mathrm{A}}}
\newcommand{\R}{\ensuremath{\mathrm{R}}}
\newcommand{\RGI}{\ensuremath{\mathrm{RGI}}}
\newcommand{\W}{\ensuremath{\mathrm{W}}}
\newcommand{\PCAC}{\ensuremath{\mathrm{PCAC}}}
\newcommand{\NSVZ}{\ensuremath{ {\mbox{\tiny \rm{NSVZ}}}}}
\newcommand{\TEK}{\ensuremath{\mathrm{TEK}}}
\newcommand{\s}{\ensuremath{\mathrm{s}}}
\newcommand{\I}{\ensuremath{\mathrm{I}}}
\newcommand{\E}{\ensuremath{\mathrm{E}}}
\newcommand{\Ep}{\ensuremath{\mathrm{E}^\prime}}

\newcommand{\Pf}{\mathrm{Pf}}

\newcommand{\cond}{\langle  {\rm Tr}  \lambda^2 \rangle}

\usepackage[font=small, labelfont=bf, textfont={small,it}]{caption}
\begin{document}

\title{Nonperturbative determination of the ${\cal N}\!=\!1$ \\ supersymmetric Yang--Mills gluino condensate at large $N$}

\author{Claudio Bonanno}
\email{claudio.bonanno@csic.es}
\affiliation{Instituto de F\'isica Te\'orica UAM-CSIC, c/ Nicol\'as Cabrera 13-15, Universidad Aut\'onoma de Madrid, Cantoblanco, E-28049 Madrid, Spain}

\author{Pietro Butti}
\email{pbutti@unizar.es}
\affiliation{Departamento de F\'isica T\'eorica, Facultad de Ciencias and\\Centro de Astropart\'iculas y F\'isica de Altas Energ\'ias (CAPA),\\ Universidad de Zaragoza, Calle Pedro Cerbuna 12, E-50009, Zaragoza, Spain}

\author{Margarita Garc\'ia P\'erez}
\email{margarita.garcia@csic.es}
\affiliation{Instituto de F\'isica Te\'orica UAM-CSIC, c/ Nicol\'as Cabrera 13-15, Universidad Aut\'onoma de Madrid, Cantoblanco, E-28049 Madrid, Spain}

\author{\\Antonio Gonz\'alez-Arroyo}
\email{antonio.gonzalez-arroyo@uam.es}
\affiliation{Instituto de F\'isica Te\'orica UAM-CSIC, c/ Nicol\'as Cabrera 13-15, Universidad Aut\'onoma de Madrid, Cantoblanco, E-28049 Madrid, Spain}
\affiliation{Departamento de F\'isica Te\'orica, Universidad Aut\'onoma de Madrid, M\'odulo 15, Cantoblanco, E-28049 Madrid, Spain}

\author{Ken-Ichi Ishikawa}
\email{ishikawa@theo.phys.sci.hiroshima-u.ac.jp}
\affiliation{Core of Research for the Energetic Universe, Graduate School of Advanced Science and Engineering, Hiroshima University, Higashi-Hiroshima, Hiroshima 739-8526, Japan}
\affiliation{Graduate School of Advanced Science and Engineering, Hiroshima University, Higashi-Hiroshima, Hiroshima 739-8526, Japan}

\author{Masanori Okawa}
\email{okawa@hiroshima-u.ac.jp}
\affiliation{Graduate School of Advanced Science and Engineering, Hiroshima University, Higashi-Hiroshima, Hiroshima 739-8526, Japan}

\date{\today}

\begin{abstract}
We present the first nonperturbative large-$N$ calculation of the $\mathcal{N}=1$  supersymmetric (SUSY) $\mathrm{SU}(N)$ Yang--Mills gluino condensate obtained by means of numerical simulations of the lattice-discretized theory, exploiting large-$N$ twisted volume reduction. We present two different determinations based, respectively, on the Banks--Casher formula and on the Gell-Mann--Oakes--Renner relation, both giving perfectly consistent results. By expressing the lattice results in the Novikov--Shifman--Vainshtein--Zakharov (NSVZ) scheme, we are able for the first time to compare numerical and analytic computations. Our most accurate determination of the Renormalization Group Invariant (RGI) gluino condensate gives $\Sigma_{\rm RGI} /\Lambda_{\rm NSVZ}^3 = [1.18\, (08)_{\rm stat}\, (12)_{\rm syst}]^3 = 1.64(33)_{\rm stat} \, (50)_{\rm syst} = 1.64(60)$, in agreement with the $N$-dependence and the value predicted by the weak coupling instanton-based approach $\Sigma_{\rm RGI} /\Lambda_{\rm NSVZ}^3 = 1$.
\end{abstract}

\maketitle

\section{Introduction}

The value of the gluino condensate in the ${\cal N} = 1$ supersymmetric (SUSY) $\SU(N)$ Yang–Mills theory has been a subject of debate since the first exact instanton-based calculations were conducted in the early 1980s~\cite{Novikov:1983ee,Rossi:1983bu,Amati:1984uz,Novikov:1985ic}. A comprehensive overview can be found in~\cite{Hollowood:1999qn}. At that time, two distinct methodologies were employed to obtain the gluino condensate. In one approach, the so-called strong-coupling (SC) instanton approach~\cite{Novikov:1983ee,Rossi:1983bu,Amati:1984uz}, the gluino condensate was derived from the one-instanton contribution to the $2N$-point function 
$\langle \Tr \lambda^2 (x_1) \cdots \Tr \lambda^2 (x_N) \rangle$, a quantity related, assuming clustering, to the object of interest  $\langle \Tr \lambda^2\rangle^N$. In contrast, the weak-coupling (WC) instanton approach incorporated additional matter fields and operated within a Higgs phase, enabling a controlled weak coupling calculation of the condensate~\cite{Novikov:1985ic}. By invoking holomorphicity, this approach allowed to decouple the additional matter fields, leading to the value of the condensate in the ${\cal N} = 1$ SUSY theory.

The two methods yielded two distinct values for the Renormalization Group Invariant (RGI) gluino condensate, exhibiting different leading $N$-dependence in the large-$N$ limit:
\be\label{eq.SigmaInst}
\Sigma_\RGI\equiv \frac{1}{(4\pi) ^2 b_0 N} \left |\cond \right|= 
\begin{cases}  
2 e  \, \Lambda^3_\NSVZ / N, \quad &\text{SC} \,,\\
\Lambda^3_\NSVZ, \quad &\text{WC} \,,
\end{cases}
\ee
where $\Lambda_\NSVZ$ refers to the $\Lambda$-parameter in the Novikov--Shifman--Vainshtein--Zakharov (NSVZ) scheme, given at-all-orders by~\cite{Novikov:1983ee,Shifman:1986zi}: 
\be\label{eq.NSVZLambda}
\Lambda_\NSVZ^3  = \frac{\mu^3}{b_0 \lambda_\NSVZ(\mu) } \exp \left ( \frac {-8 \pi^2}{\lambda_\NSVZ(\mu)} \right)\,,
\ee
with $\lambda_\NSVZ(\mu)$ denoting the renormalized 't Hooft coupling in this scheme, and $b_0=3/(4\pi)^2$, the first universal coefficient of the ${\cal N}=1$ SUSY $\beta$-function.\footnote{The $\Lambda$-parameter is defined following the standard QCD convention, trivially mapped to the usual ones in SUSY instanton calculations, see, e.g., Ref.~\cite{Armoni:2003yv}.}

An alternative approach to computing the gluino condensate is through the use of fractional instantons, namely, self-dual configurations with fractional topological charge $Q=1/N$~\cite{tHooft:1981nnx}. In contrast to standard instantons, these allow for direct saturation of the value of the gluino two-point function, circumventing the need for clustering~\cite{Cohen:1983fd,Zhitnitsky:1989ds}. However, it was not until recently that the proportionality constant between the condensate and $\Lambda^3_\NSVZ$ was computed using fractional objects. The result obtained in the WC approach has been reproduced using the fractional constituents of finite-temperature calorons~\cite{Davies:1999uw,Davies:2000nw}. In contrast, using the fractional instanton solutions~\cite{GarciaPerez:2000aiw,Gonzalez-Arroyo:2019wpu} on asymmetric tori endowed with 't Hooft's twisted boundary conditions~\cite{tHooft:1979rtg} results in a value that, for gauge group $\SU(2)$, is 2 times larger than that of WC~\cite{Anber:2022qsz}. This factor of 2 is expected to become a factor of $N$ for $\SU(N)$,\footnote{Recently, a new study~\cite{Anber:2024mco} from the same authors of~\cite{Anber:2022qsz} appeared, where a solution to this apparent discrepancy was found, resulting in an $N$-dependence in agreement with ours.} adding to the controversy surrounding the determination of the condensate.

Given the nonperturbative nature of the gluino condensate, numerical Monte Carlo simulations of the lattice-discretized theory offer a natural framework to compute this quantity from first principles. Despite a significant progress in the last decade concerning SUSY Yang--Mills lattice calculations~\cite{Giedt:2008xm,Kim:2011fw,Munster:2014bea,Munster:2014cja,Bergner:2017ytp,Ali:2018fbq,Ali:2018dnd,Ali:2019agk,Bergner:2019dim,Piemonte:2020wkm,Bergner:2024ttq} (see also Refs.~\cite{Bergner:2016sbv,Bergner:2022snd,Schaich:2022xgy} and references therein), determining the gluino condensate has proven to be a highly non-trivial numerical challenge. So far, only a few $\SU(2)$ determinations have been obtained~\cite{Giedt:2008xm,Kim:2011fw,Bergner:2019dim,Piemonte:2020wkm}, all lacking continuum limit, and all missing the matching to the NSVZ scheme, thus preventing comparison with analytic predictions.

In this paper we present the first nonperturbative computation of the gluino condensate at large-$N$ using a lattice-discretized version of the supersymmetric theory. We anticipate that our result, extrapolated to the continuum and zero-gluino-mass limit, and matched to the NSVZ scheme, will be in agreement with the WC prediction, including the $N$-dependence and the numerical coefficient of the condensate. 

The paper is organized as follows. In Sec.~\ref{sec.schemes} we examine the conventions that have led to the definition of  $\Sigma_{\rm RGI}$, as given in Eq.~\eqref{eq.SigmaInst}. In Sec.~\ref{sec.methods} we present the two different methodologies used to determine the gluino condensate on the lattice. The final results for the gluino condensate are presented in Sec.~\ref{sec.results}. All technical details concerning the lattice simulations, the determination of the condensate, and of the dynamically-generated scale $\Lambda_\NSVZ$, are discussed respectively in Apps.~\ref{app.TEK}, \ref{app.condensate}, and~\ref{app.lambda}.

\section{From the lattice to the NSVZ scheme}
\label{sec.schemes}

Before presenting our results, it is essential to establish the conventions that underpin the definition of $\Sigma_{\rm RGI}$, as given in Eq.~\eqref{eq.SigmaInst}.\footnote{A discussion along the lines presented here, but restricted to leading order perturbation theory, was given in Ref.~\cite{Armoni:2003yv}.} It should be noted that, although defined in the NSVZ scheme, $\Sigma_{\rm RGI}$, or equivalently $\cond$, is a renormalization group invariant and scheme-independent quantity~\cite{Shifman:1986zi,Hisano:1997ua,Arkani-Hamed:1997lye,Arkani-Hamed:1997qui}. The relationship between the quantity $\cond$ and the renormalized gluino two-point function in the NSVZ scheme, defined in terms of canonically-normalized gluino fields and dependent on the chosen scheme and scale, is given by~\cite{Shifman:1986zi,Hisano:1997ua}:
\be\label{eq.SigmaNSVZ}
\Sigma_\R^{(\NSVZ)}(\mu) =  \frac{N\left[1-\lambda_\NSVZ (\mu)/(8\pi^2)\right] }{\lambda_\NSVZ(\mu)} \vert \cond \vert \,. 
\ee
To obtain a generalisation of this expression for any arbitrary renormalisation scheme ``$\s$'', the starting point is the Callan--Symanzik equations, which define the $\beta$-function and the gluino mass anomalous dimension:
\beq
\beta_\s(\lambda_\s) = \frac{{\rm d} \lambda_\s}{{\rm d} \log(\mu^2)}\,, \quad
\tau_{\rm s}(\lambda_\s) = \frac{{\rm d} \log\left(m_{\R}^{(\s)}(\mu)\right)}{{\rm d} \log(\mu)}\,.
\eeq
The equation for the mass anomalous dimension can be formally integrated. The result, when combined with the renormalization group invariance of $m_{\R}^{(\s)}(\mu) \Sigma_{\R}^{(\s)}(\mu)$, can be used to define a renormalization group and scheme-independent condensate~\cite{Gasser:1982ap,Sint:1998iq,Giusti:1998wy,DellaMorte:2005kg}:
\beq\label{eq:RGI_def}
\Sigma_{\rm RGI} &= &  {\cal A} \, \Sigma_\R^{(\s)}(\mu)  \, 
\left [2 b_0\lambda_\s(\mu)\right]^\frac{d_0}{2b_0} \nonumber \\
\times&\exp&\left[\int_0^{\lambda_\s(\mu)} dx
\left(\frac{\tau_\s(x)}{2 \beta_\s(x)}-\frac{1}{ x}\right)\right]\,,
\label{eq:condensate}
\eeq
where the coefficients $d_0$, $b_0$ and $b_1$ represent the universal terms in the asymptotic expansions of the $\beta$ and $\tau$-functions. For ${\cal N}=1$ supersymmetric Yang-Mills theory, these coefficients take the values $(4\pi)^2b_0=3$, $(4\pi)^4b_1=6$, and $d_0=2 b_0$.
The normalization factor ${\cal A}$ can be readily determined by employing the relationship between the exact $\beta$-function and the mass anomalous dimension in the NSVZ scheme~\cite{Jack:1997pa}:
\beq
\frac{\tau_\NSVZ(x)}{2 \beta_\NSVZ(x)}&=& \frac{1}{x (1 - b_1 x/ b_0) } \,.
\eeq
Upon insertion of this expression into Eq.~\eqref{eq:condensate}, a comparison to Eqs.~\eqref{eq.SigmaInst} and~\eqref{eq.SigmaNSVZ} yields the result ${\cal A} =  8\pi^2/ (9 N^2)$.

The final ingredient necessary to match the lattice- and instanton-based determinations of the condensate is the RGI $\Lambda$-parameter in a conventionally defined scheme such as $\MS$, which can be reliably determined on the lattice. This dynamically-generated scale is defined by formally integrating the Callan--Symanzik equation defining the $\beta$-function, leading to:
\beq
\Lambda_\s&=&\mu\left[b_0
\lambda_\s(\mu)\right]^\frac{-b_1}{2{b_0}^2}\exp\left(\frac{-1}{2b_0\lambda_\s(\mu)}\right) \nonumber\\
\times&\exp&\left[-\int_0^{\lambda_\s(\mu)} dx
\left(\frac{1}{2 \beta_s(x)}+\frac{1}{2 b_0 x^2}-\frac{b_1}{2b_0^2x}\right)\right]\,.
\label{eq.Lambda}
\eeq
A one-loop calculation enables the matching of the NSVZ and $\MS$ schemes, as expressed by the equation: $\Lambda_\NSVZ = e^{-1/18} \Lambda_{\MS}$~\cite{Finnell:1995dr}.

When all the elements are combined, the result, at next-to-leading order, is (see App.~\ref{app.condensate}):
\be
\Sigma_{\RGI} = {\cal A}\, 2 b_0 \lambda_{\MS}(\mu) \, \Big [ 1  +  \frac{d_1^{(\MS)}\!-\!2b_1}{2b_0} \lambda_{\MS}(\mu) \Big] \Sigma_\R^{(\MS)}(\mu) \,,
\ee
where one can rely on the two-loop relation between the renormalized coupling and the $\Lambda$-parameter, i.e.:
\beq
2 b_0\lambda_{\MS}(\mu) = -\frac{1}{\log(y)} - \frac{b_1}{2 b_0^2} \frac{ \log\left[\log(1/y^2)\right]}{\log^2(y)}\, ,
\eeq
with $y \equiv \Lambda_{\MS}/\mu$. Using these expressions, we will be able to determine the RGI condensate in terms of the lattice determination of $\Sigma_\R^{(\MS)}(\mu)$.

\section{Methods}
\label{sec.methods}

Our lattice-based extraction of the renormalized gluino condensate employs two distinct methodologies. The first is a Banks--Casher-based approach, which relates the spectral density of the Dirac operator to the condensate. The second is based on a Gell-Mann--Oakes--Renner-like relation that is applicable in SUSY Yang--Mills theory when endowed with a gluino mass SUSY-breaking term (see below). The two aforementioned methods, employed in SUSY Yang--Mills for the first time in this study, have recently been successfully employed by the authors to determine the quark condensate in large-$N$ Yang--Mills theories~\cite{Bonanno:2023ypf}. It should be also noted that the spectral method has been extensively used in the QCD literature~\cite{Giusti:2008vb,Luscher:2010ik,Cichy:2013gja,Cichy:2013rra,Cichy:2015jra,Bonanno:2019xhg,Athenodorou:2022aay,Bonanno:2023xkg,Bonanno:2023ypf}.

The Banks--Casher (BC) equation establishes a relationship between the condensate in the massless-gluino limit and the value of the spectral density at the origin.  In their seminal paper, Leutwyler and Smilga, proposed the following expression for the specific case of ${\cal N }= 1$ supersymmetric Yang-Mills theory~\cite{Leutwyler:1992yt}:
\beq\label{eq:banks_casher}
\frac{\Sigma}{2\pi} = \lim_{\lambda\to 0}\lim_{m\to0}\lim_{V\to\infty} \rho(\lambda,m),
\eeq
where $i \lambda + m$ stands for a generic eigenvalue of the massive Dirac operator $\slashed{D} + m$. The additional factor of two with respect to the usual BC relation has to do with the Majorana nature of gluinos~\cite{Leutwyler:1992yt}.

Giusti and L\"uscher demonstrated that a more convenient quantity for determining the condensate on the lattice is the mode number~\cite{Giusti:2008vb}. This is defined in relation to the spectral density, as follows:
\beq\label{eq:modenumber_def}
\braket{\nu(M, m)}  =  V \int_{-\lambda_M}^{-\lambda_M} d \lambda \, \rho(\lambda,m) \,,
\eeq
with $\lambda_M= \sqrt{M^2-m^2}$. The integration of Eq.~\eqref{eq:banks_casher} between $-\lambda_M$ and $\lambda_M$  is straightforward, leading in the chiral and thermodynamic limit to:
\beq\label{eq:sigma_eff}
\Sigma  \equiv  \lim_{m\to0}\lim_{V\to\infty}  \frac{ \pi \nu(M, m)}{ 4 V \lambda_{M}}\,.
\eeq
The advantage of this formulation is that the mode number is a renormalization group invariant quantity~\cite{Giusti:2008vb}, which means that the renormalization properties of $\Sigma$ are fully dictated by those of $\lambda_M$.  Further technical details about the practical implementation of the Giusti--L\"uscher method and its related lattice renormalization procedure can be found in App.~\ref{app.condensate}.

The gluino condensate can also be determined from a Gell-Mann--Oakes--Renner-like (GMOR) relation. As is well known, the lattice formulation explicitly breaks supersymmetry. However, Kaplan~\cite{Kaplan:1983sk} and Curci and Veneziano~\cite{Curci:1986sm} demonstrated that the ${\cal N}=1$ SUSY Yang--Mills theory could be recovered as the continuum limit of a lattice Yang--Mills theory coupled to a massless adjoint Majorana fermion. In the continuum and chiral limits of such a theory, supersymmetry is recovered as an accidental symmetry. The chiral behaviour of the non-singlet adjoint-pion, an unphysical particle not present in the SUSY spectrum, can be used to determine the point where SUSY gets restored. The mass of this adjoint-pion is related, at leading order in the chiral expansion, to the gluino mass by a relation analogous to the Gell-Mann--Oakes--Renner one in QCD~\cite{Munster:2014bea,Munster:2014cja}:
\beq
m_\pi^2 = 2\frac{\Sigma_\R^{(\s)}}{F^2_\pi} m_\R^{(\s)}\, ,
\eeq
where by $F_\pi$ we denote the adjoint-pion decay constant in the massless gluino limit. 
By employing this relationship, one can derive a determination of the renormalized gluino condensate, provided that the mass of the adjoint-pion and its decay constant, as well as the renormalized gluino mass, are known on the lattice.
In App.~\ref{app.condensate} we provide all the details concerning the lattice computation of $F_\pi$ and $m_\R^{(\s)}$. 

It should be noted that the methods described above allow for the determination of $\Sigma_\R^{(\s)}/Z_\S^{(\s)}$ and  $\Sigma_\R^{(\s)}/Z_\P^{(\s)}$. A nonperturbative determination of $Z_\S$ and  $Z_\P$ represents a highly non-trivial challenge on its own, that falls beyond the scope of this paper. Consequently, we have opted to rely on a nonperturbative determination of $Z_\P/Z_\S$ based on the spectral methods described in Refs.~\cite{Giusti:2008vb,Luscher:2010ik,Bonanno:2019xhg,Athenodorou:2022aay,Bonanno:2023xkg,Bonanno:2023ypf}, combined with the two-loop determination of $Z_\S^{(\MS)}(\mu)$ provided in Ref.~\cite{Skouroupathis:2007jd}. In order to keep into account the perturbative renormalization and the two-loop truncation of Eqs.~\eqref{eq:RGI_def} and~\eqref{eq.Lambda}, a $30\% $ systematic error has been added to our final results for the condensate. The size of the systematic error has been chosen on the basis of the typically-observed mismatch between 2-loop and nonperturbative results for renormalization constants in Yang--Mills theories~\cite{Bali:2013kia,Perez:2020vbn,Castagnini:2015ejr,Bonanno:2023ypf} (for explicit examples, see Tab.~6 of~\cite{Perez:2020vbn}, where a $\sim 10\%$ between perturbative and nonperturbative determinations of $Z_\A$ is reported, and Tabs.~6 and 7 of~\cite{Bali:2013kia} where $\sim 16\%$ and $\sim 37\%$ are reported for, respectively, $Z_\S$ and $Z_\P$ at $N=3$, $b=0.334$). This will be our dominant source of uncertainty.

\section{Results}
\label{sec.results}

The gluino condensate has been determined using the Monte Carlo ensembles generated in Ref.~\cite{Butti:2022sgy}. The reader is referred to this paper and to App.~\ref{app.TEK} for all the technical details concerning the simulations. These were performed using the so-called large-$N$ twisted volume reduction~\cite{Gonzalez-Arroyo:1982hyq,Gonzalez-Arroyo:2010omx,Gonzalez-Arroyo:2010omx}, which was generalized to include the case of adjoint fermions in Ref.~\cite{Gonzalez-Arroyo:2013bta}. The relationship between space-time and colour degrees of freedom that emerges in the large-$N$ limit due to Eguchi--Kawai equivalence~\cite{Eguchi:1982nm} enables the large-$N$ theory to be simulated as a model of $\SU(N)$ matrices on a lattice with a single space-time point and twisted boundary conditions~\cite{tHooft:1979rtg}. 
This equivalence has been successfully employed to investigate numerous instances of Yang--Mills theories in the large-$N$ limit~\cite{Gonzalez-Arroyo:2012euf,GarciaPerez:2014azn,Perez:2020vbn,Bonanno:2023ypf}, as well as several non-supersymmetric theories with Dirac fermions in the adjoint representation~\cite{GarciaPerez:2015rda}. This technique was applied in Ref.~\cite{Butti:2022sgy} to simulate ${\cal N}=1$ supersymmetric Yang-Mills theory, generating configurations with dynamical massive gluinos at several values of the inverse bare 't Hooft coupling ($b=1/\lambda_{\rm L}=0.34$, 0.345, 0.35 are used here), and for several values of $N$ (169, 289 and 361). In accordance with the strategy proposed by the DESY--Jena--Regensburg--M\"unster collaboration~\cite{Ali:2019agk,Ali:2018fbq,Ali:2018dnd}, we employed Wilson fermions and adopted the well-known reweighting method to address the computation of the Pfaffian, which is not positive definite for this particular fermion discretization. This entailed incorporating the Pfaffian sign in the observable and sampling the Monte Carlo configurations with its modulus. In~\cite{Butti:2022sgy} we explicitly computed the sign of the Pfaffian on the ensembles used for this study, and found it was always positive, thus no reweighting was needed. In~\cite{Butti:2022sgy} we also determined the lattice spacing in physical units, a prerequisite for relating any lattice-based measurement to a common physical standard. To this end, we employed the hadronic reference scale $\sqrt{8t_0}$, based on the renormalization group flow of the gauge action density, which has emerged as the standard for lattice computations~\cite{Luscher:2010iy}, and which will be used subsequently in the estimation of the $\Lambda$-parameter. 

\begin{figure}[!t]
\centering
\includegraphics[scale=0.5]{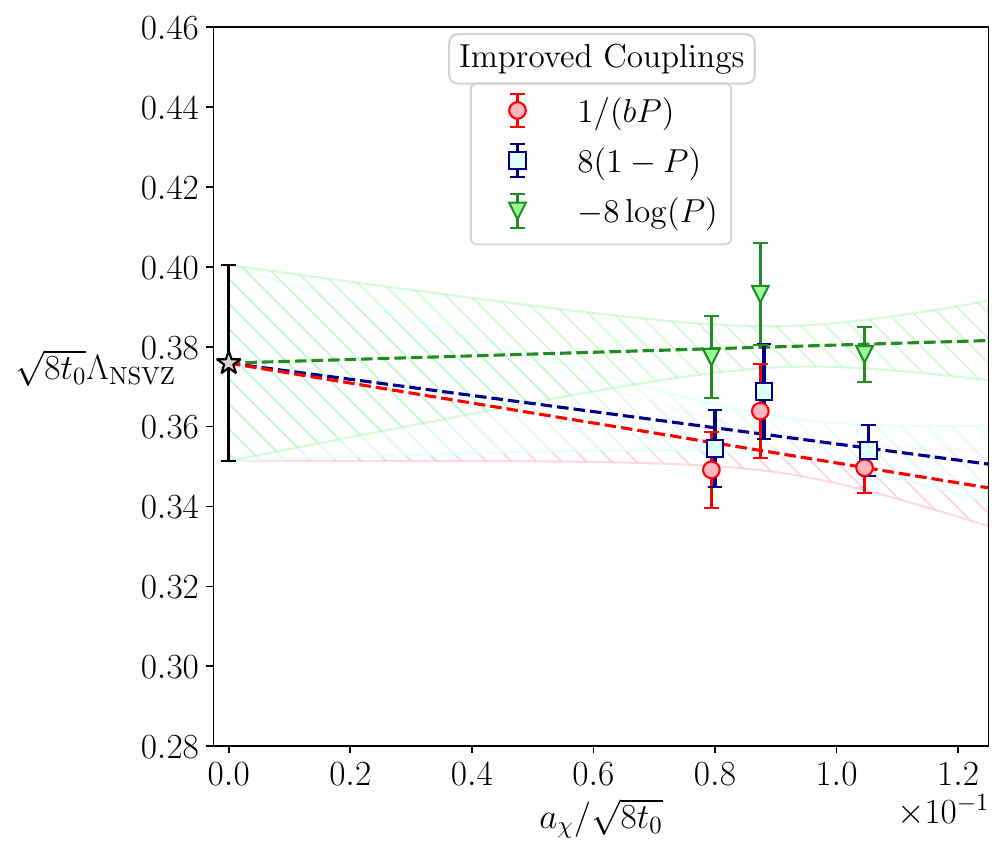}
\caption{Determination of $\sqrt{8 t_0} \Lambda_{\NSVZ}$ (star) from asymptotic scaling with different improved couplings assuming $\mathcal{O}(a_\chi)$ artefacts. Dashed lines and shaded areas represent the joint continuum extrapolation best fit obtained imposing a common continuum limit.}
\label{fig:asymptotic_scaling}
\end{figure}

Our determination of the dynamically-generated scale of the theory, $\Lambda$, relies on the use of asymptotic scaling, which has proven to be highly reliable in the case at hand. By making use of the defining equation~\eqref{eq.Lambda}, we determined the value of the $\Lambda$-parameter relative to the hadronic scale $\sqrt{8t_0}$ through the two-loop expression:
\beq
\begin{aligned}
\sqrt{8t_0}\Lambda_{\s} &= \lim_{a_\chi\to 0} 
\frac{\sqrt{8t_0}}{a_{\chi}} \exp\{-f(\lambda_{\s})\} \, , \\
f(x) &= \frac{1}{2 b_0}\left[\frac{1}{x} + \frac{b_1}{b_0} \log(b_0 x)\right]\, ,
\end{aligned}
\eeq
where the momentum scale has been set to $\mu=1/a_{\chi}$, with $a_{\chi}$ the lattice spacing in the massless-gluino limit obtained in~\cite{Butti:2022sgy}. It is widely acknowledged that the use of the naive bare coupling, $\lambda_{\rm L}=1/b$, is not expected to exhibit good scaling properties. However, it is possible to accelerate the convergence of the perturbative expansion by defining appropriate improved couplings.  In the literature, there are several standard choices based on the use of the average plaquette $P$. In this study, we considered three possibilities: $\lambda_t^{(\I)} = 1/(bP)$, $\lambda_t^{(\E)} = 8(1-P)$, and $\lambda_t^{(\Ep)} = -8\,\log(P)$. Each of these definitions introduces a different lattice scheme, with an associated determination of the $\Lambda$-parameter. The matching of all these definitions to the $\MS$ scheme is known; details on the matching prescriptions are provided in App.~\ref{app.lambda}. Figure~\ref{fig:asymptotic_scaling} shows the continuum extrapolation leading to the determination of $\sqrt{8t_0}\Lambda_{\NSVZ}$. The values of $a_\chi/\sqrt{8t_0}$ shown in the figure correspond to the chiral (zero-gluino-mass) limit, as obtained in Ref.~\cite{Butti:2022sgy}. As previously noted, asymptotic scaling works remarkably well and leads to:
\be \sqrt{8 t_0} \Lambda_{\NSVZ} = 0.376(25) \, .
\label{eq:lambda_res}
\ee
For the sake of completeness we also quote: $\sqrt{8 t_0} \Lambda_{\MS} = 0.397(26)$.

\begin{figure}[!t]
\centering
\includegraphics[scale=0.5]{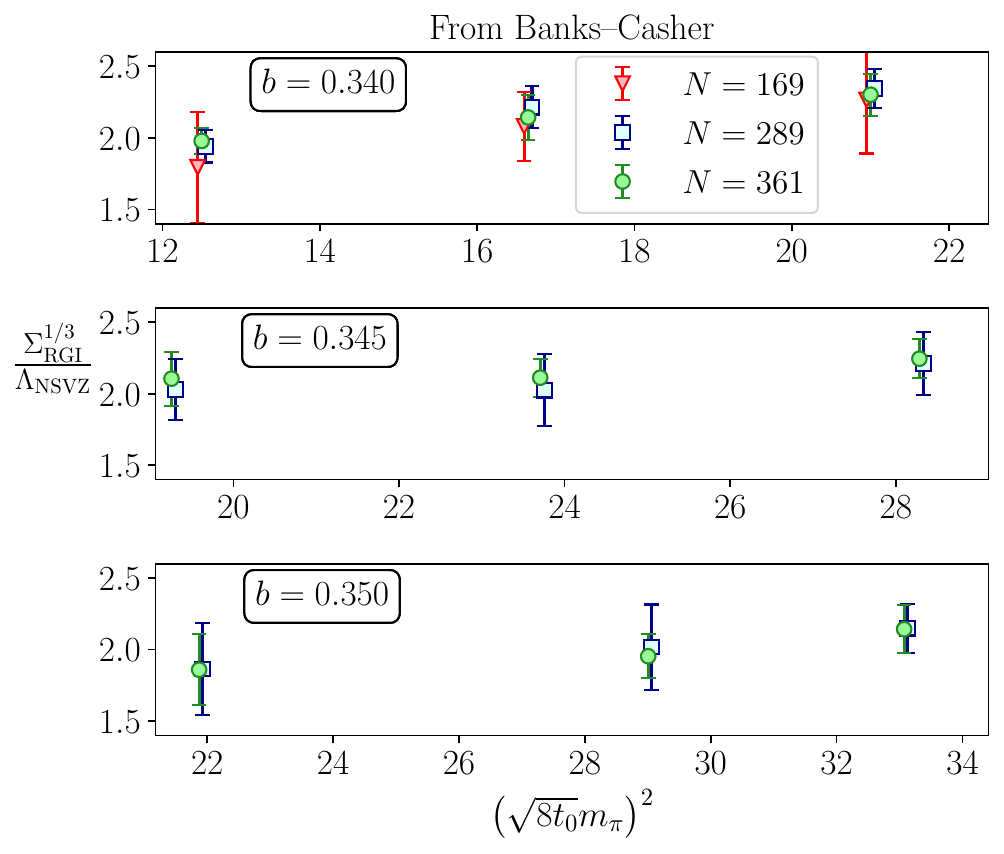}
\includegraphics[scale=0.5]{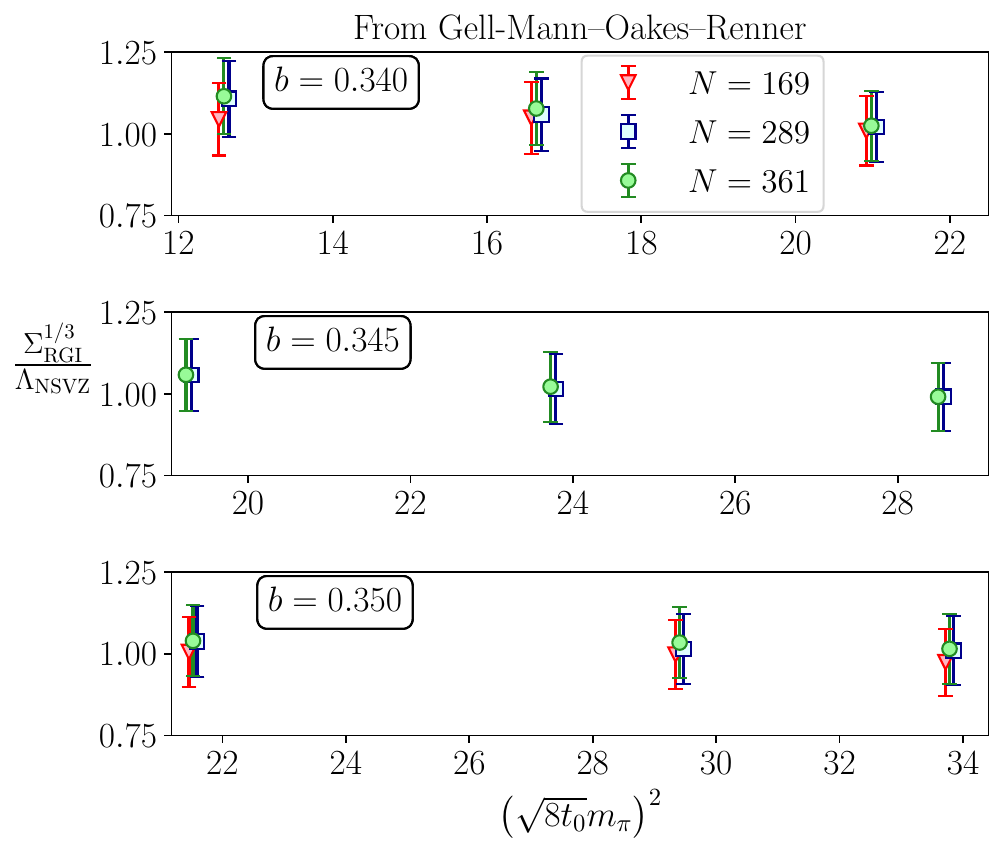}
\caption{Behavior of the gluino condensate $\Sigma_{\RGI}^{1/3}/\Lambda_{\NSVZ}$ as a function of $N$ for several values of the adjoint-pion mass $m_\pi$ and the bare 't Hooft coupling $1/b$. The x-axis refers to the values of $\sqrt{8t_0} m_\pi$ determined for $N=361$. Points slightly shifted horizontally for readability.}
\label{fig:gluino_cond_comp_N}
\end{figure}

We now proceed to the calculation of the gluino condensate. In this section, we will merely present the final results. Further details on the extraction of the condensate using either the BC or GMOR methods can be found in App.~\ref{app.condensate}. A summary of all the intermediate numerical results used to extract the condensate in the supersymmetric limit is provided in Tab.~\ref{tab:summary_results} in App.~\ref{app.condensate}. It should be noted that in order to determine the condensate using the GMOR method, the value of the adjoint-pion decay constant in the SUSY (chiral-continuum) limit was required. This was calculated, resulting in a value:
\be 
\frac{F_\pi}{N\Lambda_{\NSVZ}}=0.092(14)\, .
\ee

A significant outcome of our analysis is the leading $N$-dependence of the condensate, when written in units of $\Lambda_\NSVZ$, c.f.~Eq.~\eqref{eq.SigmaInst}. The three available values of $N$, namely $N$ = 169, 289, and 361, were analyzed using both the BC and the GMOR methods. In the BC approach, we were able to reliably determine the condensate only for volumes satisfying $m_\pi \ell_{\rm eff} = m_\pi (a\sqrt{N}) \, \gtrsim \, 10$. The outcomes at finite gluino mass are presented as functions of the pion mass squared in Fig.~\ref{fig:gluino_cond_comp_N}. There is no observed dependence on the number of colours. This result, in full accordance with the WC instanton-based approach, is clearly in contradiction with the $1/N$ dependence predicted by the SC determination.

\begin{figure}[!t]
\centering
\includegraphics[scale=0.5]{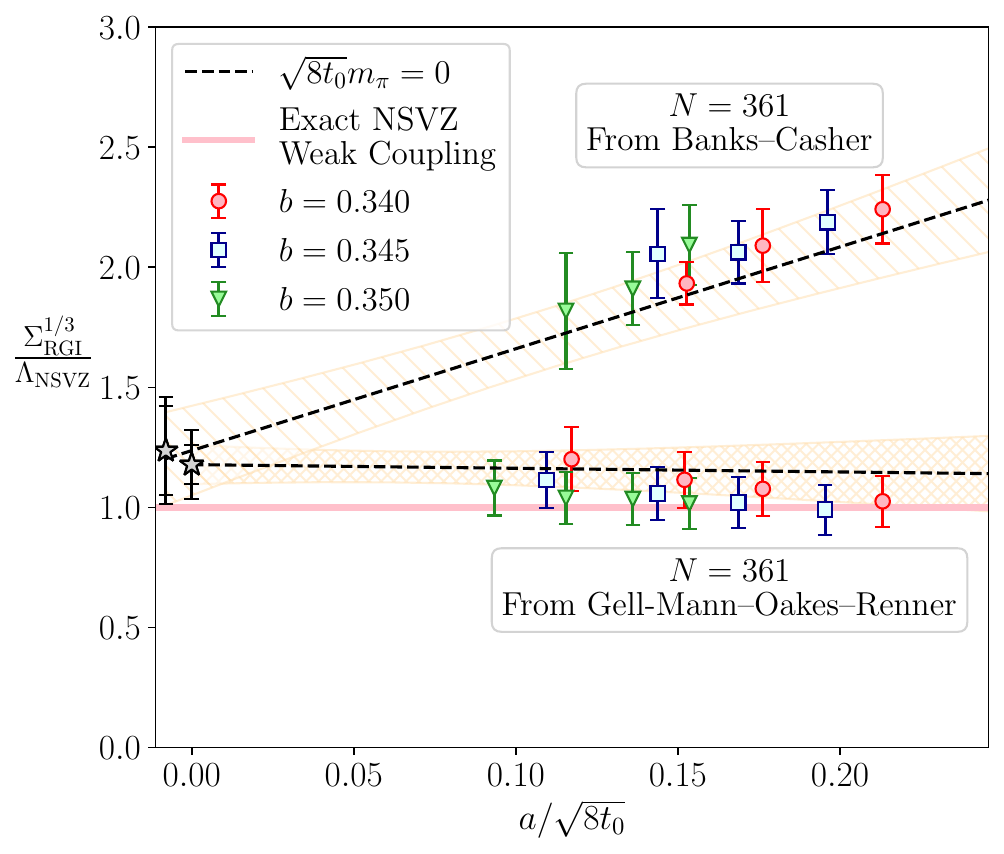}
\caption{Chiral-continuum extrapolations of the gluino condensate (stars) obtained from the Banks--Casher and from the Gell-Mann--Oakes--Renner relations, assuming $\mathcal{O}(a)$ and $\mathcal{O}(m_\pi^2)$ corrections. Dashed lines and shaded areas represent the best fit with the coefficient of the $\mathcal{O}(m_\pi^2)$ correction set to zero}.
\label{fig:gluino_cond}
\end{figure}

\begin{table}[!t]
\small
\begin{center}
\begin{tabular}{|c|c|c|c|c|c|}
\hline
&&&&&\\[-1em]
$b$ & $\kappa$ & $\frac{a}{\sqrt{8 t_0}}$ & $m_{\pi}\sqrt{8 t_0}$ & \makecell{$\Sigma_{\RGI}^{1/3}/\Lambda_{\NSVZ}$\\(BC)} & \makecell{$\Sigma_{\RGI}^{1/3}/\Lambda_{\NSVZ}$\\(GMOR)} \\
\hline
\multirow{4}{*}{0.340} & 0.1850 & 0.2132 & 4.58 & 2.24(14) & 1.02(11)\\
& 0.1890 & 0.1762 & 4.08 & 2.09(15) & 1.08(11)\\
& 0.1910 & 0.1521 & 3.55 & 1.93(8) & 1.11(12)\\
& 0.1930 & 0.1172 & 2.24 & - & 1.20(13)\\
\hline
\multirow{4}{*}{0.345} & 0.1800 & 0.1954 & 5.34 & 2.19(13) &  0.99(10)\\
& 0.1840 & 0.1686 & 4.87 & 2.06(13) & 1.02(11)\\
& 0.1868 & 0.1438 & 4.39 & 2.06(18) & 1.06(11)\\
& 0.1896 & 0.1095 & 3.23 & - & 1.11(12)\\
\hline
\multirow{4}{*}{0.350} & 0.1800 & 0.1535 & 5.81 & 2.09(16) & 1.01(11)\\
& 0.1825 & 0.1361 & 5.42 & 1.91(15)& 1.03(11)\\
& 0.1850 & 0.1155 & 4.64 & 1.82(24)& 1.04(11)\\
& 0.1875 & 0.0934 & 3.07 & - & 1.08(11)\\
\hline
\end{tabular}
\end{center}
\caption{Summary of the numerical results reported in Fig.~\ref{fig:gluino_cond}, and used to extrapolate the gluino condensate towards the SUSY (i.e., chiral-continuum) limit.}
\label{tab:final_res_fig3}
\end{table}

Given the excellent agreement among determinations for different values of $N$, the final numbers for the gluino condensate were obtained through a joint chiral and continuum extrapolation of the results at $N=361$, collected in Tab.~\ref{tab:final_res_fig3} and obtained at finite gluino mass and lattice spacing as described in Apps.~\ref{app.condensateBC} and~\ref{app.condensateGMOR}. The scale is set by fixing $\sqrt{8 t_0} \Lambda_{\NSVZ}$ to the value given in Eq.~\eqref{eq:lambda_res}.

In Fig.~\ref{fig:gluino_cond}, displaying the values reported in Tab.~\ref{tab:final_res_fig3} of $\Sigma_\RGI^{1/3}/\Lambda_{\NSVZ}$ as a function of the lattice spacing in units of $\sqrt{8t_0}$, we perform a simultaneous chiral and continuum extrapolation of the gluino condensate assuming $\mathcal{O}(8t_0m_\pi^2)$ and $\mathcal{O}(a/\sqrt{8t_0})$ corrections, as expected from chiral perturbation theory for Wilson fermions. The dashed lines and shaded areas represent
the best fit with the coefficient of the  $\mathcal{O}(8t_0m_\pi^2)$ correction set to zero.
The results are compared to those predicted by the WC instanton-based approach, represented by the horizontal line indicating $\Sigma_\RGI^{1/3}/\Lambda_\NSVZ=1$. As can be observed, the BC and the GMOR methods yield results which are perfectly in accordance when extrapolated to the supersymmetric limit:
\beq
\begin{aligned}
\frac{\Sigma_{\RGI}}{\Lambda_{\NSVZ}^3} &= [1.18(08)_{\rm stat} \, (12)_{\rm syst}]^3\\
\\[-1em]
&= 1.64(33)_{\rm stat} \, (50)_{\rm syst}
\end{aligned}
\quad \text{(GMOR)}
\eeq
\beq
\begin{aligned}
\frac{\Sigma_{\RGI}}{\Lambda_{\NSVZ}^3} &= [1.24(19)_{\rm stat}(12)_{\rm syst}]^3&\\
\\[-1em]
&= 1.89(87)_{\rm stat} \, (55)_{\rm syst}
\end{aligned}
\quad \text{(BC)}
\eeq
Furthermore, these numbers are in good agreement with the WC determination. Although the errors in our calculation are significant, it is crucial to highlight that this is the first time a numerical determination of this kind has been carried out in the existing literature. Previous calculations of the condensate in four-dimensional supersymmetric Yang-Mills theory~\cite{Giedt:2008xm,Kim:2011fw,Bergner:2019dim,Piemonte:2020wkm} were limited to a single lattice spacing and gauge group $\SU(2)$, and were never matched to the NSVZ scheme.

\section{Conclusions}
\label{sec.conclusions}

In this paper we have presented the first nonperturbative lattice calculation of the gluino condensate in ${\cal N}=1$ supersymmetric Yang-Mills theory in the large-$N$ limit. We have exploited large-$N$ volume independence, and adopted two independent strategies, giving perfectly agreeing results. After extrapolation to the supersymmetric limit and matching with the NSVZ scheme, our results confirm the $N$-dependence predicted by semi-classical instanton methods based on the so-called weak coupling (WC) approach~\cite{Novikov:1985ic,Davies:1999uw,Davies:2000nw}. Our most accurate determination of the condensate in units of the dynamically-generated scale of the theory is: $\Sigma_{\RGI} /\Lambda_{\NSVZ}^3 = [1.18 \, (08)_{\rm stat}\,(12)_{\rm syst}]^3 = 1.64(33)_{\rm stat} \, (50)_{\rm syst} = 1.64(60)$, consistent with the expected result from WC. The error budget in our result is largely influenced by the systematic error resulting from the absence of a nonperturbative determination of the renormalisation constant  $Z_{\S}$. While addressing this issue would be a valuable avenue for future research, it would also necessitate the commitment to a long-term project that is beyond the scope of the current study.

\begin{acknowledgments}
It is a pleasure to thank José L.~F.~Barbón, Georg Bergner, Gregorio~Herdo\'iza, Nikolai Husung, Carlos Pena and Erich Poppitz for useful discussions. This work is partially supported by the Spanish Research Agency (Agencia Estatal de Investigaci\'on) through the grant IFT Centro de Excelencia Severo Ochoa CEX2020-001007-S, funded by MCIN/AEI/10.13039 /501100011033, and by grant PID2021-127526NB-I00, funded by MCIN/AEI/10.13039/ 501100011033 and by “ERDF A way of making Europe”. We also acknowledge support from the project H2020-MSCAITN-2018-813942 (EuroPLEx) and the EU Horizon 2020 research and innovation programme, STRONG-2020 project, under grant agreement No 824093. P.~B.~is supported by Grant PGC2022-126078NB-C21 funded by MCIN/AEI/ 10.13039/501100011033 and “ERDF A way of making Europe”. P.~B.~also acknowledges support by Grant DGA-FSE grant 2020-E21-17R Aragon Government and the European Union - NextGenerationEU Recovery and Resilience Program on ‘Astrofísica y Física de Altas Energías’ CEFCA-CAPA-ITAINNOVA. K.-I.~I.~is supported in part by MEXT as "Feasibility studies for the next-generation computing infrastructure". M.~O.~is supported by JSPS KAKENHI Grant Number 21K03576. Numerical calculations have been performed on the \texttt{Finisterrae~III} cluster at CESGA (Centro de Supercomputaci\'on de Galicia). We have also used computational resources of Oakbridge-CX at the University of Tokyo through the HPCI System Research Project (Project ID: hp230021 and hp220011).
\end{acknowledgments}

\appendix

\section{The TEK lattice formulation of $\mathcal{N}=1$ large-$N$ $\SU(N)$ SUSY Yang--Mills}
\label{app.TEK}

Our entire lattice setup, including the gauge configuration ensembles, come from the one explained and adopted in Ref.~\cite{Butti:2022sgy}, which we briefly summarize here.

The Twisted Eguchi--Kawai (TEK) model is a $d=4$-matrix model with partition function given by:
\beq
\mathcal{Z}_{\TEK} = \int [dU] \Pf\left\{C D_{\W}^{(\TEK)}[U]\right\}e^{-S_{\TEK}[U]},
\eeq
with the charge-conjugation operator $C$ satisfying $\gamma_\mu^{\rm t} C = - \gamma_\mu C$ and $C^{\rm t}=-C$.

The gauge action, expressed in terms of the gauge link $U_\mu \in \SU(N)$ in the fundamental representation, reads:
\beq
S_{\TEK}[U] = bN\sum_{\mu\ne\nu}\Tr\left\{\mathrm{I} - z_{\mu\nu}^{*}U_\mu U_\nu U_\mu^\dagger U_\nu^\dagger\right\},
\eeq
where $z_{\mu\nu}=e^{i 2\pi n_{\mu\nu}/N}$ is the twist factor, with $\sqrt{N}\in\mathbb{N}$, $n_{\mu\nu}=\sqrt{N} k(N)$ for $\nu>\mu$ and $n_{\mu\nu}=-n_{\nu\mu}$ (we used $k=5,5,7$ for $N=169,289,361$). Instead, the adjoint Wilson--Dirac operator reads:
\beq
\begin{aligned}
&D_{\W}^{(\TEK)}[U] = \frac{1}{2 \kappa}\mathrm{I} \,\,\, + \\& - \frac{1}{2}\sum_{\mu}\bigg[\,\,\,(\mathrm{I}-\gamma_\mu)U_\mu^{(\rm adj)} + \left.(\mathrm{I}+\gamma_\mu)\left(U_\mu^{(\rm adj)}\right)^\dagger \right],
\end{aligned}
\eeq
with $\kappa$ the Wilson hopping parameter, and $U_\mu^{(\rm adj)}$ denoting the $d=4$ $\SU(N)$ gauge fields in the adjoint representation.

Since the Pfaffian of the lattice Wilson--Dirac operator is not guaranteed to be positive, we follow the approach proposed by the DESY--Jena--Regensburg--M\"unster collaboration~\cite{Ali:2019agk,Ali:2018fbq,Ali:2018dnd}, namely, we resort to sign-quenched simulations, meaning that the sign of the Pfaffian is moved to the observable via standard reweighting:
\beq
\braket{\mathcal{O}} = \frac{\left\langle \mathcal{O} \, \mathrm{sign}\left[\Pf\left\{C D_{\W}^{(\TEK)}[U]\right\}\right]\right\rangle_{\rm sq}}{\left\langle\mathrm{sign}\left[\Pf\left\{C D_{\W}^{(\TEK)}[U]\right\}\right]\right\rangle_{\rm sq}}.
\eeq
The sign-quenched (sq) expectation value is calculated sampling the following functional integral using a standard Rational Hybrid Monte Carlo (RHMC) algorithm (see Ref.~\cite{Butti:2022sgy}):
\beq
\mathcal{Z}_{\TEK}^{(\rm sq)} = \int [dU] \left\vert \Pf\left\{C D_{\W}^{(\TEK)}[U]\right\} \right\vert e^{-S_{\TEK}[U]}.
\eeq

\begin{table*}
\small
\begin{center}
\begin{tabular}{|c|c|c|c|c|c|c|c|c|c|c|}
\hline
&&&&&&&&&&\\[-1em]
$b$ & $\kappa$ & $a m_{\pi}$ & $a m_{\PCAC}$ & \makecell{$a^3 \Sigma_{\R}/(Z_\P N^2)$\\(Banks--Casher)} & $\sqrt{8 t_1}/a$ & $Z_\P/Z_\S$ & \makecell{$P$\\(Plaquette)} & $a F_\pi/(Z_\A N)$  & $\kappa_c$ & $\sqrt{8t_1}/a_\chi$\\
\hline
\hline
\multirow{4}{*}{0.340} & 0.1850 & 0.977(4) & 0.2018(14) & 0.00378(69)  & 2.883(58)   & 0.642(2)  & 0.541414(99) & 0.1758(16) & \multirow{4}{*}{0.19359(5)} & \multirow{4}{*}{5.88(11)}
\\
& 0.1890 & 0.719(4) & 0.1083(7) & 0.00213(43)  & 3.488(88)   & 0.560(5)  & 0.549479(88) & 0.1317(10) & &
\\
& 0.1910 & 0.540(5)  & 0.0609(9)  & 0.00129(17)  & 4.024(58)   & 0.500(5)  & 0.55419(10) & 0.1006(17) & &
\\
& 0.1930 & 0.263(13) & 0.0138(6)  & - & 5.244(175) & 0.410(12) & 0.560145(74) & 0.0422(13) & &
\\\hline
\hline
\multirow{4}{*}{0.345} & 0.1800 & 1.043(6) & 0.2352(32) & 0.00266(43) &  3.134(93) & 0.683(4) & 0.55072(10) & 0.16841(92) &  \multirow{4}{*}{0.19095(6)} & \multirow{4}{*}{7.03(23)}
\\
& 0.1840 & 0.821(5) &  0.1507(11) & 0.00163(28) & 3.645(98) & 0.625(6) & 0.556355(81) & 0.13412(87) & &
\\
& 0.1868 & 0.631(6) & 0.0913(18) & 0.00118(31) &  4.274(55) & 0.565(8) & 0.561253(71) & 0.1067(13) & &
\\
& 0.1896 & 0.353(5) &  0.0289(5) & - &  5.614(201) & 0.464(10) & 0.567106(91) & 0.0581(12) & &
\\
\hline
\hline
\multirow{4}{*}{0.350} & 0.1800 & 0.883(7) &  0.1826(16) & 0.00126(28) & 4.003(109) & 0.663(5) & 0.564610(70) & 0.13560(94) & \multirow{4}{*}{0.18857(5)} & \multirow{4}{*}{7.74(21)}
\\
& 0.1825 & 0.733(6) &  0.1242(15) & 0.00073(17) & 4.516(64) & 0.629(8) & 0.568020(55) & 0.11299(95) & &
\\
& 0.1850 & 0.540(6) &  0.0712(10) & 0.00046(18) & 5.323(65) & 0.552(8) & 0.571769(42) & 0.0839(13) & &
\\
& 0.1875 & 0.293(6) &  0.0195(4) & - &  6.582(287) & 0.472(10) & 0.576334(91) & 0.0422(11) & &
\\
\hline
\end{tabular}
\end{center}
\caption{Summary of obtained numerical results for $N=361$, with the only exceptions of $Z_\P/Z_\S$, which was instead calculated for $N=289$. The quantity $t_1$ is another hadronic reference scale which was used to set the scale in intermediate calculation. It is related to the more commonly-employed reference scale $t_0$, which we used to display all our final results in the main text, by the relation $\sqrt{t_0/t_1} \simeq 1.627$~\cite{Butti:2022sgy}. The results for the critical $\kappa_c$, the scale setting, the pion and PCAC masses, and the plaquette, come from Ref.~\cite{Butti:2022sgy}.}
\label{tab:summary_results}
\end{table*}

The sign of the Pfaffian on the ensembles used for this study was subjected to explicit analysis in the study presented in Ref.~\cite{Butti:2022sgy}. In light of the results presented in Refs.~\cite{Bergner:2015adz,Wuilloud:2010wxi,Bergner:2011wf,Bergner:2011zp,Piemonte:2015phd}, the sign in question can be determined by counting the number of negative real eigenvalues of $D_{\W}$. It is anticipated that, for heavy gluino masses, there will be no flips in the sign of the Pfaffian. Consequently, the analysis focused on the lightest adjoint fermion masses at each value of $b$ and $N$. No negative-real eigenvalues were identified in the spectrum. Furthermore, this absence was confirmed for a number of heavier fermion masses. Therefore, we concluded that the sign of the Pfaffian is consistently positive for the employed model parameters, thereby simplifying the reweighting method by validating the distribution using the absolute value of the Pfaffian.

\section{Extracting the gluino condensate from the lattice}
\label{app.condensate}

In this Appendix we describe the extraction of the gluino condensate from lattice gauge configurations using the two methodologies described in Sec.~\ref{sec.methods}: the spectral method and the GMOR relation. We also describe how to relate the bare quantity obtained from these methods to the RGI condensate. A summary of all intermediate numerical results for the largest value of $N=361$, along with the results of~\cite{Butti:2022sgy} employed in this study, is presented in Tab.~\ref{tab:summary_results}.

\subsection{Spectral method (Banks--Casher)}
\label{app.condensateBC}

The Giusti--L\"uscher method is a well-established numerical strategy to compute fermion condensates from lattice simulations based on the low-lying spectrum of the Dirac operator and on the Banks--Casher relation~\cite{Giusti:2008vb,Luscher:2010ik,Bonanno:2019xhg,Athenodorou:2022aay,Bonanno:2023xkg,Bonanno:2023ypf}. We summarize it in the following.

We numerically solved the Hermitian eigenvalue problem
\beq\label{eq:eigval_problem}
\left(\gamma_5 D_{\rm W}^{(\TEK)}\right) u_\lambda = \lambda \, u_\lambda, \quad \qquad \lambda \in \mathbb{R},
\eeq
for the first 100 low-lying eigenvalues using the \texttt{ARPACK} library in order to compute the mode number $\braket{\nu(M)}=\braket{\# \vert \lambda \vert \le M}$ as a function of the threshold mass $M$.

\begin{figure}[!t]
\centering
\includegraphics[scale=0.25]{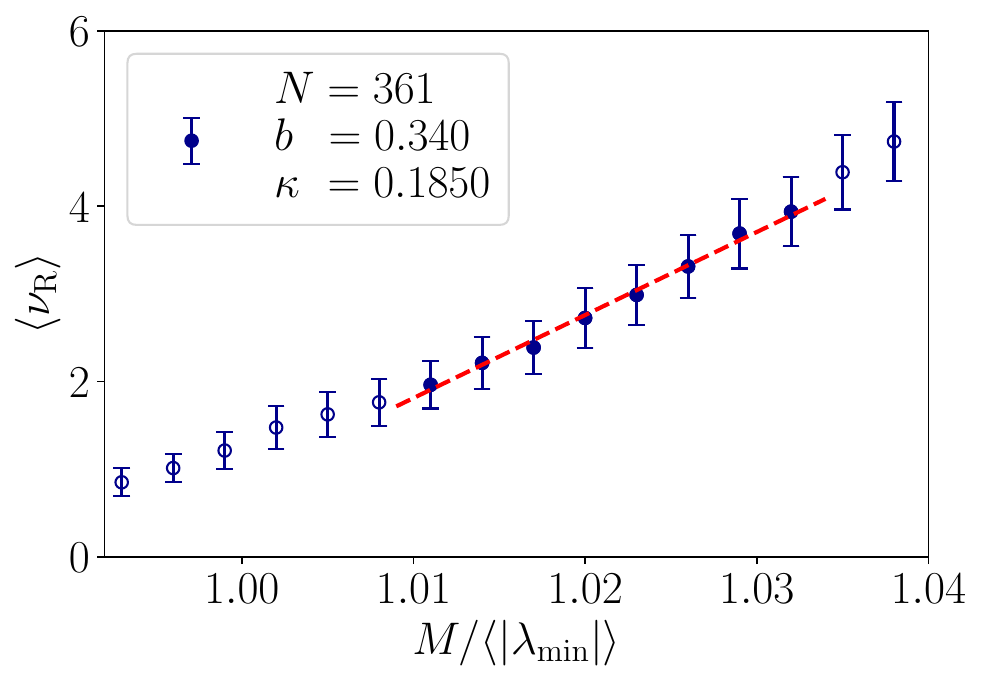}
\includegraphics[scale=0.25]{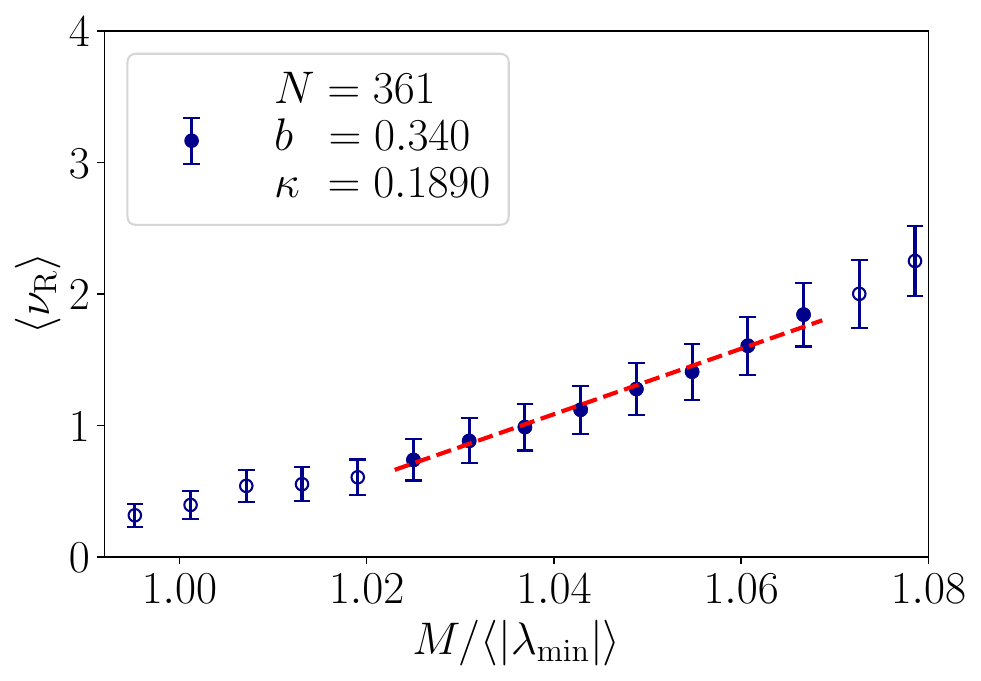}
\includegraphics[scale=0.25]{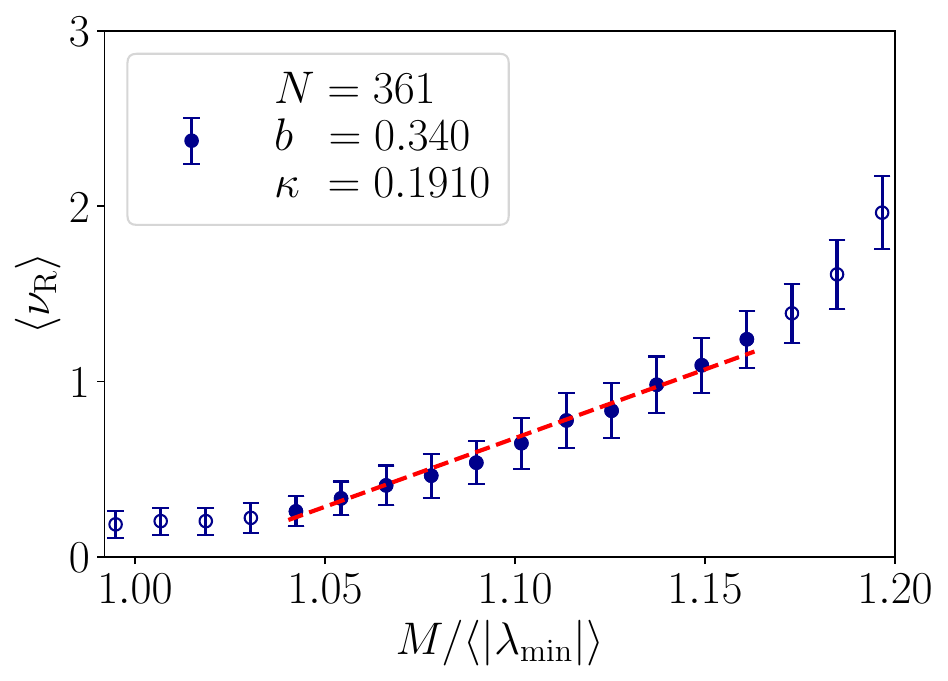}
\includegraphics[scale=0.25]{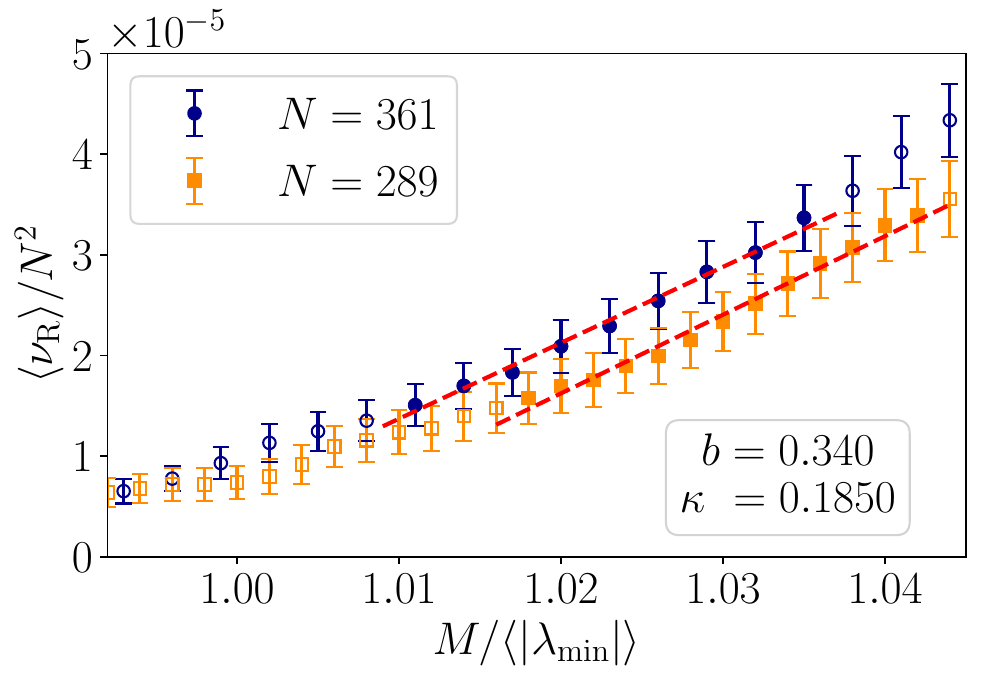}
\caption{Examples of calculation of mode number slope for $b=0.34$, $N=361$ and three values of the Wilson hopping parameter $\kappa=0.185,0.189,0.191$. In the bottom right plot we compare two different values of $N$ for $b=0.34$ and $\kappa=0.185$.}
\label{fig:mode_number_fit}
\end{figure}

Given that $M$ renormalizes as an eigenvalue ($\lambda = Z_\P \lambda_\R$), the Giusti--L\"uscher method allows to determine the renormalized condensate, up to the renormalization constant $Z_\P$, as follows:
\beq\label{eq:GL}
\frac{\Sigma_{\R}}{Z_\P} = \frac{\pi}{4V} \sqrt{1-\left(\frac{m_{\R}}{M_{\R}}\right)^2} \, S,
\eeq
with $V=a^4$ the one-point lattice volume and $a$ the lattice spacing, and with
\beq
S = \frac{\partial \braket{\nu(M)}}{\partial M}
\eeq
the slope of the mode number as a function of the threshold mass $M$, calculated for $M$ close to $M = \braket{\vert \lambda_{\rm min} \vert }$ (the average minimum eigenvalue extrapolated to the infinite $N$ limit). Note that $\Sigma_\R/Z_\P$ is RGI, thus we do not need to specify any renormalization scheme/renormalization scale. Examples of mode number best fits to extract the slope are shown in Fig.~\ref{fig:mode_number_fit}. 

Given that the Banks--Casher relation just gives the leading order term of the chiral expansion of the spectral density in powers of $m$ and $\lambda$, the linearity of the mode number as a function of $M$ is only valid up to $O(M^2)$ corrections. To avoid contaminations from higher-order terms, we chose the the fit range according to the same criteria suggested in the original paper~\cite{Giusti:2008vb}, which were also applied in other works addressing the spectral determination of fermion condensates, see, e.g., Refs.~\cite{Engel:2014cka,Bonanno:2023ypf,Bonanno:2023xkg}.

The lower bound of the fit range was chosen so as to stay close to the threshold $M/\braket{\vert \lambda_{\min} \vert }=1$. For Wilson fermions, the lower end of the spectrum is known to deviate from the expected continuum form~\cite{Giusti:2008vb}, therefore we actually started the fit range in the linear region a bit above the threshold. Then, we tried several linear best fits increasing the upper bound of the fit range until a good reduced chi-square ($p$-value larger than 5\%) could be obtained. We also verified that including/excluding one further point from the lower/upper bound of the fit range did not change significantly the obtained slope, i.e., that the contamination coming from higher-order terms in $M$ is negligible. Thus, systematic errors related to the choice of the fit range are negligible compared to our statistical ones.

Given the renormalization property of the so-called ``subtracted definition'' of the bare gluino mass:
\beq
a m_{\rm sub} = \frac{1}{2\kappa}-\frac{1}{2\kappa_c} = a Z_{\S} m_{\R},
\eeq
with $\kappa_c$ the critical $\kappa$ obtained in~\cite{Butti:2022sgy}, in order to calculate the factor $m_{\R}/M_{\R} = (Z_\P/Z_\S)(m_{\rm sub}/M)$ appearing in Eq.~\eqref{eq:GL} we need the RGI ratio $Z_\P/Z_\S$.

\begin{figure}[!t]
\centering
\includegraphics[scale=0.41]{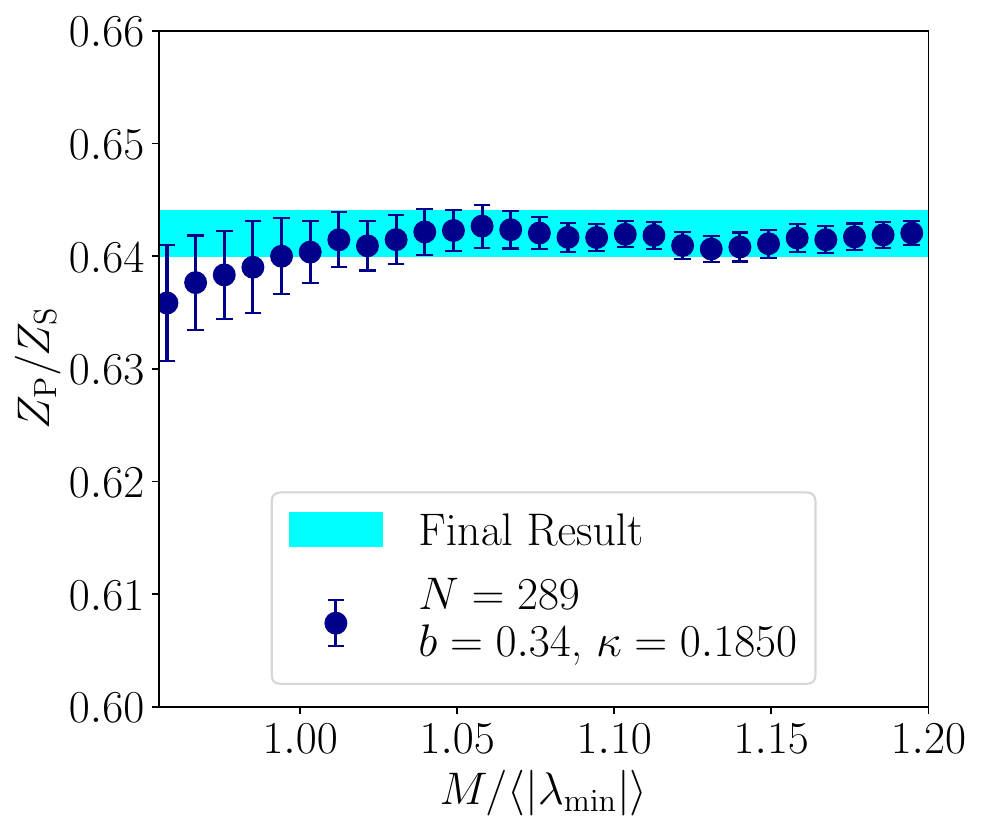}
\caption{Example of nonperturbative determination of $Z_\P/Z_\S$ using the spectral method, Eq.~\eqref{eq:ZPZS_spec}, for $N=289$, $b=0.34$ and $\kappa=0.185$. The shaded area represents our final result for the ratio $Z_\P/Z_\S$.}
\label{fig:ZPZS}
\end{figure}

This quantity can be computed from spectral methods using the eigenvectors obtained from the same eigenvalue problem in Eq.~\eqref{eq:eigval_problem}~\cite{Giusti:2008vb,Luscher:2010ik,Cichy:2013gja,Cichy:2013rra,Cichy:2015jra,Bonanno:2019xhg,Athenodorou:2022aay,Bonanno:2023xkg,Bonanno:2023ypf} as follows:

\beq\label{eq:ZPZS_spec}
\left(\frac{Z_\P}{Z_\S}\right)^2 = \frac{\braket{s_\P(M)}}{\braket{\nu(M)}},\\
\nonumber\\
\nonumber\\
s_\P(M) \equiv \sum_{\vert\lambda\vert,\vert\lambda^{\prime}\vert \le M} \vert u_\lambda^\dagger \gamma_5 u_{\lambda^\prime}\vert^2,
\eeq
where the ratio in Eq.~\eqref{eq:ZPZS_spec} is expected to exhibit a plateau as a function of $M$ for sufficiently large values of $M$. The expected behavior is verified from actual lattice data, see Fig.~\ref{fig:ZPZS}. Since $Z_\P/Z_\S$ is a ratio of spectral sums, and since it is determined by the plateau at larger values of $M$, while finite-size effects mostly affect the smallest eigenvalues, finite-volume (i.e., finite-$N$) corrections for this quantity are expected to be largely negligible.

Tab.~\ref{tab:summary_results} collects the values of $a^3 \Sigma_{\R}(a, m_\pi)/(Z_\P N^2)$, obtained applying this method for $N=361$, and the values of $Z_\P/Z_\S(a, m_\pi) $, determined nonperturbatively for $N=289$. These are converted in physical units through:
\beq
\frac{\Sigma_\R (a, m_\pi) }{Z_\S N^2\Lambda^3_{\NSVZ}} &= & \left(\frac{\sqrt{8 t_0}}{a}\right)^3 \times \frac{a^3 \Sigma_{\R}(a, m_\pi) }{Z_\P N^2} \times \frac{Z_\P}{Z_\S}(a, m_\pi)  \times  \nonumber \\ && \left (\sqrt{8t_0} \Lambda_{\NSVZ} \right)^{-3}\, .
\eeq
After using the perturbatively determined value of $Z_{\S}$, determined as described in App.~\ref{app.ren}, and the procedure to pass from the renormalized to the RGI condensate, again described in App.~\ref{app.ren}, the determinations reported in Tab.~\ref{tab:final_res_fig3} are obtained.

\subsection{Pion mass method (Gell-Mann--Oakes--Renner)}
\label{app.condensateGMOR}

Although softly-broken SUSY Yang--Mills has just a single massive fermion flavor as its dynamical content, it is possible to consider, in the rigorous framework of partially-quenched chiral perturbation theory, the chiral behavior of the adjoint pion, an unphysical particle whose mass is related to the gluino one by a Gell-Mann--Oakes--Renner-like relation, just as in QCD~\cite{Munster:2014bea,Munster:2014cja}. In this context, one finds:
\beq
m_\pi^2 = 2\frac{\Sigma_\R}{F^2_\pi}m_{\R}
\eeq
where $F_\pi$ refers to the decay constant of the unphysical adjoint pion, while $\Sigma_\R$ is the physical gluino condensate we wish to calculate. Clearly, both quantities must be understood as computed in the chiral (i.e., SUSY-restoring) limit $m_\R \to 0$.

From the adjoint-pion mass correlators computed in~\cite{Butti:2022sgy} it is possible to extract both $m_\pi$ and $F_\pi/Z_\A$. Once one gets rid of $Z_\A$ and isolates $F_\pi$, from the knowledge of $m_\pi$ as a function of the bare gluino mass it is possible to obtain the renormalized gluino condensate, up to the renormalization constant $Z_S$, as:
\beq
\frac{\Sigma_\R}{Z_\S} = \frac{m_\pi^2 F_\pi^2}{2 m_{\rm sub}}.
\label{eq:sig_zp}
\eeq

To obtain $F_\pi/Z_\A$, we used standard techniques relying on the calculation of the pion-vacuum matrix element of the temporal component of the axial vector current:
\beq
\frac{F_\pi}{N Z_\A} = \frac{1}{\sqrt{2} N m_\pi} \langle 0 \vert A_4(x=0) \vert \pi(\vec{p}=0) \rangle,
\eeq
where the normalization was chosen on the basis of standard counting arguments, leading to $F_\pi\sim\mathcal{O}(N)$ for adjoint fermions. The details of the Generalized EigenValue Problem (GEVP) solved to find the largest-overlapping interpolating pion operator for twisted-reduced models can be found in the appendices of Refs.~\cite{Perez:2020vbn} and~\cite{Butti:2022sgy}. Tab.~\ref{tab:summary_results} collects the values of $a F_\pi/(N Z_A)$ obtained for $N=361$.

\begin{figure}[!t]
\centering
\includegraphics[scale=0.5]{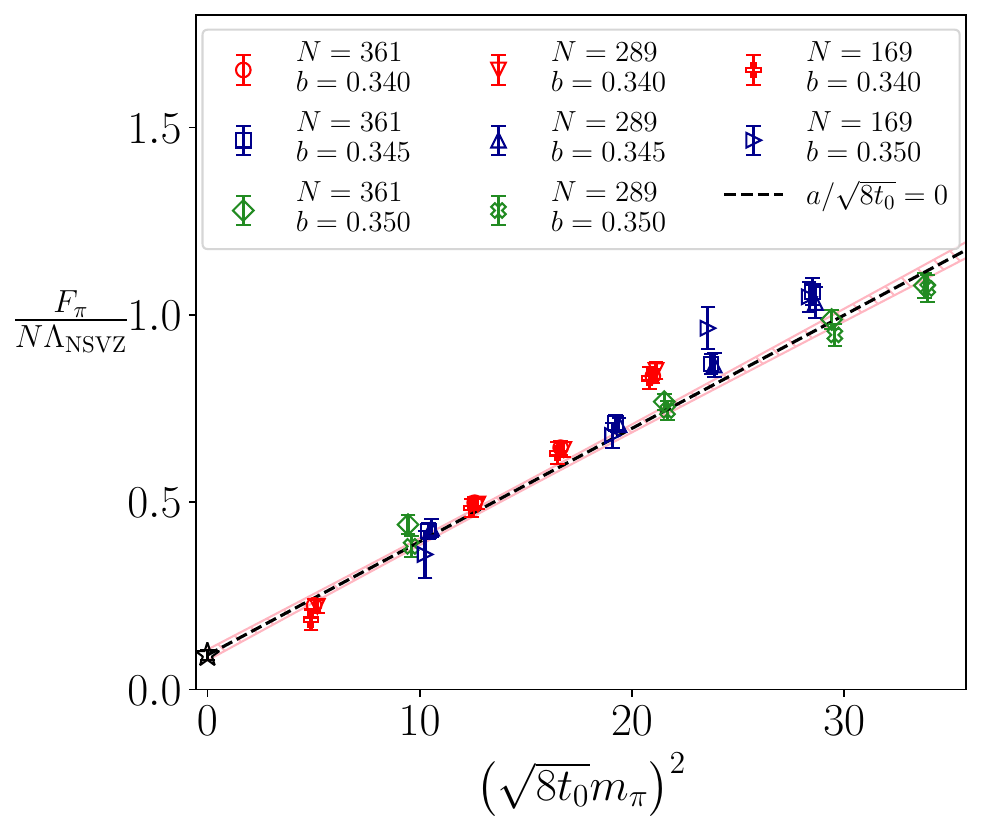}
\caption{Chiral-continuum extrapolation (star) of the $N=361$ results for the adjoint-pion decay constant $F_\pi/(N\Lambda_{\NSVZ})$. For comparison we also show results obtained for $N=169$ and 289. The dashed line and the shaded area represent the chiral extrapolation best fit at the zero-lattice-spacing point. Best fit yields a reduced $\tilde{\chi}^2\simeq 14.4/9$.}
\label{fig:Fpi}
\end{figure}

Since $F_\pi$ needs renormalization, it is useful to recall that from the axial Ward Identity it is possible to extract a different definition of the bare quark mass, the so-called PCAC (Partially-Conserved Axial Current) mass:
\beq
m_\PCAC = \frac{Z_\P}{Z_\A} m_\R.
\eeq
Therefore, using the spectral determination of $Z_\P/Z_\S$, and the PCAC mass determinations of~\cite{Butti:2022sgy}, the adjoint-pion decay constant, for each value of $N$, $b$ and gluino mass, is obtained via:
\beq
\frac{F_\pi}{N} = \left(\frac{F_\pi}{N Z_\A}\right) \times \left(\frac{m_{\rm sub}}{m_{\PCAC}}\right) \times \left(\frac{Z_\P}{Z_\S}\right),
\eeq
as in this combination the factors of $Z_\P/(Z_\S Z_\A)$ exactly cancel out. 

The values of the decay constant obtained in this way (normalized to the value of $ \Lambda_{\NSVZ}$) are displayed as a function of the adjoint-pion mass squared, in units of $8t_0$, in Fig.~\ref{fig:Fpi}. A joint chiral and continuum extrapolation of all the results obtained for the largest $N = 361$ is then performed. Assuming $\mathcal{O}(a)$ and $\mathcal{O}(m_\pi^2)$ corrections, we fit to a functional form:
\be
\frac{\sqrt{8t_0} F_\pi (a, m_\pi) } {N \sqrt{8t_0} \Lambda_{\NSVZ}} = \frac{F_\pi}{N  \Lambda_{\NSVZ}} + c_1 \frac{a}{\sqrt{8 t_0}} + c_2 \, 8 t_0 m_\pi^2
\,,\ee
where $\sqrt{8t_0} \Lambda_{\NSVZ}$ is set to the value in the SUSY limit, given by Eq.~\eqref{eq:lambda_res}, and the values of $\sqrt{8t_0} F_\pi (a, m_\pi)$ can be determined in a straightforward way from the data given in Tab.~\ref{tab:summary_results}.
Higher-order corrections in the chiral expansion appear to be negligible within our numerical precision, as adding a $O(m_\pi^4)$ power to the fit function does not change the obtained result for $F_\pi$ in the SUSY limit. The dashed line and the shaded area in Fig.~\ref{fig:Fpi} represent the pion mass dependence obtained by setting $c_1=0$.  We obtain a value for the  adjoint-pion decay constant in the chiral and continuum limit  given by: $F_\pi / (N \Lambda_{\NSVZ}) = 0.092(14)$.

Using this value of $F_\pi$ and the values of the adjoint pion and PCAC masses determined in Ref.~\cite{Butti:2022sgy} (compiled in Tab.~\ref{tab:summary_results} for $N=361$) it is straightforward to ascertain the values of $\Sigma_{\R}/Z_{\S}$ through the application of Eq.~\eqref{eq:sig_zp}. To be precise, we determine:
\beq
\frac{\Sigma_\R(a, m_\pi) }{Z_\S N^2\Lambda^3_{\NSVZ}} &=& \left(\frac{\sqrt{8t_0}}{a}\right) \times \frac{(a m_\pi)^2}{ 2 (a m_{\rm sub})} \times \left(\frac{F_\pi}{\sqrt{N}\Lambda_{\NSVZ}}\right)^2 \times\nonumber \\ && \frac{1}{ \sqrt{8t_0} \Lambda_{\NSVZ}} .
\eeq
The outcomes obtained for the largest $N=361$, multiplied by the perturbatively determined value of $Z_{\S}$, and passed to the RGI definition, c.f.~App.~\ref{app.ren}, are summarised in Tab.~\ref{tab:final_res_fig3}.

\subsection{Renormalization and conversion to RGI definition}
\label{app.ren}

The methods described so far allow to determine the bare gluino condensates $\Sigma_\R/Z_\P$ and $\Sigma_\R/Z_\S$. Given that we are also able to obtain nonperturbative determinations of $Z_\P/Z_\S$, in the end we are just left with two independent determinations of, say, $\Sigma_\R/Z_\S$. A nonperturbative determination of $Z_\S$ is not available at present, therefore we have to rely on 2-loop perturbation theory to estimate $Z^{(\MS)}_\S(\mu)$. In terms of the renormalized coupling at scale $\mu=1/a$, the renormalization constant $Z_{\S}$ in the $\MS$ scheme at two-loop order reads~\cite{Skouroupathis:2007jd}:
\beq
\begin{aligned}
Z_{\S}^{(\MS)}\left(\mu=\frac{1}{a}, \lambda_{\MS}\right) & = 1 - \frac{12.9524104(1)}{(4\pi)^2} \lambda_{\MS}\\
\\[-1em]
\\[-1em]
& - \frac{60.68(10)}{(4\pi)^4} \lambda_{\MS}^2  + \mathcal{O}\left(\lambda_{\MS}^3\right)
\end{aligned}
\eeq
The renormalized coupling in the $\MS$ scheme at $\mu=1/a$ was computed using improved couplings (see App.~\ref{app.lambda}), and no significant dependence on the choice of the particular improved lattice scheme was observed in the obtained results.

With these ingredients we are able to determine, up to the systematic error involved in the perturbative determination of $Z_\S$, the value of $\Sigma_{\R}^{(\MS)} (\mu=1/a)$. As explained in the main text, the renormalized condensate $\Sigma_{\R}^{(\s)}(\mu)$, determined in any arbitrary scheme $(s)$ at an arbitrary renormalization scale $\mu$, can be converted to a scheme and scale-independent RGI quantity as follows~\cite{Gasser:1982ap,Sint:1998iq,Giusti:1998wy,DellaMorte:2005kg}:
\beq
\Sigma_{\rm RGI} &= &  {\cal A} \, \Sigma_\R^{(\s)}(\mu)  \, 
\left [2 b_0\lambda_\s(\mu)\right]^\frac{d_0}{2b_0} \nonumber \\
\nonumber\\
\times&\exp&\left[\int_0^{\lambda_\s(\mu)} dx
\left(\frac{\tau_\s(x)}{2 \beta_\s(x)}-\frac{1}{ x}\right)\right]\,,
\eeq
with $\mathcal{A}=8\pi^2/(9N^2)$ a normalization factor chosen to match the analytic calculation conventions of the NSVZ determination. At two-loop order, expanding $\beta$ and $\tau$, this relation simplifies as:
\beq
\begin{aligned}
\Sigma_\RGI &= \mathcal{A} \, 2b_0 \Sigma_{\R}^{(\s)}(\mu)  \, \lambda_{\rm s}(\mu) \\ 
&\nonumber\\
&\times \left[ 1  + \lambda_{\rm s}(\mu) \left(\frac{d_1^{(\s)}}{2b_0} -  \frac{b_1}{b_0}\right) 
\right],
\end{aligned}
\eeq
where $b_0$, $b_1$ and $d_0=2b_0$ are universal, and $d_1^{(\s)}$ is scheme-dependent. In the $\MS$ scheme at scale $\mu=1/a$, one obtains:
\beq\label{eq:rgiconv1}
\begin{aligned}
\Sigma_{\RGI} &= \mathcal{A} \, 2 b_0 \Sigma_{\R}^{(\MS)}(\mu=1/a) \, \, \lambda_{\MS}(\mu=1/a) \\
&\\
& \,\, \times \left[ 1 + \frac{d_1^{(\MS)}\!-\!2b_1}{2b_0} \lambda_{\MS}(\mu=1/a) \right].
\end{aligned}
\eeq
with $d_1^{(\MS)} = 32/(4\pi)^4$~\cite{Vermaseren:1997fq}.

Given the order of truncation we are working at, we rely on 2-loop perturbation theory also to express the 't Hooft coupling in terms of the dynamically-generated scale of the theory~\cite{Giusti:1998wy}, which we are able to reliably compute from the lattice (see App.~\ref{app.lambda}):
\beq\label{eq:rgiconv2}
\begin{aligned}
& 2 b_0\lambda_{\MS}(\mu=1/a) =\\
\\[-1em]
&-\frac{1}{\log(a \Lambda_{\MS})} - \frac{b_1}{2 b_0^2} \frac{ \log\left[-2\log(a \Lambda_{\MS})\right]}{\log^2(a \Lambda_{\MS})} \, .
\end{aligned}
\eeq
We have checked that truncating Eqs.~\eqref{eq:rgiconv1}--\eqref{eq:rgiconv2} at 1-loop order leads to a change in the final results for the condensate which is well below the 30\% systematic error we have assigned to the use of the perturbative formula for the renormalization factor $Z_{\S}$.

\section{The dynamically-generated scale $\Lambda_{\NSVZ}$ from the lattice}
\label{app.lambda}

The dynamically-generated scale $\Lambda_\NSVZ$ can be obtained from the one of the $\MS$ scheme, which is more amenable to be computed on the lattice, via: $\Lambda_{\NSVZ} = e^{-1/18} \Lambda_{\MS}$~\cite{Finnell:1995dr}.

We computed $\Lambda_{\MS}$ via asymptotic scaling:
\beq
\begin{aligned}
\sqrt{8t_0}\Lambda_{\MS} &= \lim_{a_\chi\to 0} 
\frac{\sqrt{8t_0}}{a_{\chi}} \exp\{-f(\lambda_{\MS})\} \, ,
\\
f(x) &= \frac{1}{2 b_0}\left[\frac{1}{x} + \frac{b_1}{b_0} \log(b_0 x)\right]\, ,
\end{aligned}
\eeq
with $a_\chi$ the lattice spacing extrapolated to the massless-gluino limit, computed in~\cite{Butti:2022sgy}.

Instead of matching directly the bare lattice 't Hooft coupling $\lambda_{\rm L} =1/b$ to the $\MS$ scheme, it is better first to pass through an intermediate scheme defined by the so-called improved couplings $\lambda_t^{(\s)}$, which are related to the bare one by:
\beq
\lambda_{\rm L} = \lambda_{t}^{(\s)} - 2 b_0 \left(\lambda_{t}^{(\s)}\right)^2 \log(\Lambda_{\s}/\Lambda_{\rm L}),
\eeq
with $\Lambda_{\rm L}$ the $\Lambda$-parameter related to $\lambda_{\rm L}$.

In this work we have considered 3 different improved couplings: $\lambda_t^{(\I)} = 1/(bP)$, $\lambda_t^{(\rm E)} = 8(1-P)$, and $\lambda_t^{(\Ep)} = -8\,\log(P)$, where $P$ is the averaged clover-plaquette $P \equiv \frac{1}{6N}\sum_{\mu>\nu} \braket{\Tr\, \mathrm{Clover}_{\mu\nu} }$. All of these improved couplings define its own renormalization scheme, which can be matched to the bare lattice one via:
\beq
\log\left( \frac{\Lambda_{\rm L}}{\Lambda_{\I} } \right) &= -\dfrac{w_1}{2b_0} &= \log\left(\frac{\Lambda_{\rm L}}{\Lambda_{\MS}} \times 2.7373 \right) , \\
\log\left( \frac{\Lambda_{\rm L}}{\Lambda_{\rm E} } \right) &= -\dfrac{w_2}{2b_0 w_1} &= \log\left(\frac{\Lambda_{\rm L}}{\Lambda_{\MS}} \times 29.005\right),  \\
\log\left( \frac{\Lambda_{\rm L}}{\Lambda_{\Ep} } \right) &= -\dfrac{a_1}{2b_0} &= \log\left(\frac{\Lambda_{\rm L}}{\Lambda_{\MS}} \times 5.60\right).
\eeq
Here $w_1$ and $w_2$~\cite{Weisz:1980pu,Perez:2017jyq} are the first two coefficients of the perturbative expansion of the plaquette $P$:
\beq
w_1 &=& \frac {1}{8}, \\
w_2 &=& 0.0051069297 - N_{\rm f} \, 0.0013858405(1),\\
a_1 &=& \frac{w_2}{w_1} + \frac{w_1}{2},
\eeq
with $N_{\rm f}= 1/2$. Finally, using the known relation between $\Lambda_{\rm L}$ and $\Lambda_{\MS}$ in a theory with adjoint fermions~\cite{Weisz:1980pu}:
\beq
\frac{\Lambda_{\MS}}{\Lambda_{\rm L}} = 73.46674161161081 \, ,
\eeq
and the values of the plaquette determined in Ref.~\cite{Butti:2022sgy} (c.f.~Tab.~\ref{tab:summary_results} for $N=361$) we have all the necessary ingredients to pass from improved lattice couplings to the $\MS$ one.

The result of applying this exercise to determine $\Lambda_{\NSVZ}$ through the three different improved couplings is presented in Fig.~\ref{fig:asymptotic_scaling}, in sec.~\ref{sec.results}.  Asymptotic scaling works remarkably well leading to a value: $\sqrt{8 t_0} \Lambda_{\NSVZ} = 0.376(25)$.

\end{document}